# Experimentally Exploring the Interatomic Potential in a Ferroelectric Crystal via Optimal Ultrafast Lattice Control

Blake S. Dastrup[1,a], Jacob R. Hall[1,a], Jeremy A. Johnson[1,b]

[1]Brigham Young University, Department of Chemistry and Biochemistry, Provo, UT 84602

We present a direct comparison between resonant terahertz (THz) and nonresonant impulsive stimulated Raman scattering (ISRS) excitation of phonon-polaritons in ferroelectric lithium niobate. THz excitation offers advantages of selectively driving only the forward propagating phonon-polariton mode to exceedingly high amplitudes, without complications due to nonlinear processes at the high 800 nm pump fluences used in Raman excitation. At peak-to-peak THz electric field strengths exceeding 1 MV/cm, the ferroelectric lattice is driven into the anharmonic regime, allowing experimental determination of the shape of the potential energy surface.

Ultrafast control over a crystalline lattice is of interest in developing a basic understanding of how light can influence material properties, as well as potential practical applications such as the development of ultrafast switches and a variety of optoelectronic applications [1-4]. Strong THz radiation, with frequencies resonant to the modes of interest, has been promoted as a preferred means of lattice control [1] over two common routes: displacive excitation [5,6] and nonresonant Raman excitation of vibrations [7-9]. In typical displacive-type excitation, energy is deposited into the material's electronic subsystem, transiently distorting the lattice. Such excitation may coherently drive many modes simultaneously, but potentially leaves large amounts of (unwanted) incoherent thermal energy. Nonresonant ultrafast Raman excitation, termed impulsive stimulated Raman scattering (ISRS) [7-9], can coherently excite lattice modes without the excess thermal energy, but it is an inefficient process and thus extremely high-fluence laser pulses are required to drive sizeable amplitude vibrations. In the attempt to selectively excite large amplitude vibrations for lattice control, in either displacive excitation or ISRS, high pump pulse fluences end up initiating multiple nonlinear optical processes that interfere with the excitation and even lead to sample damage. On the other hand, relatively low-photon-energy THz radiation has the potential to resonantly and efficiently excite vibrational modes to large amplitudes without the same danger of permanent sample damage. This potential advantage of THz excitation has been discussed the last few years as high-field THz sources have become more common, and yet to our knowledge no direct comparison between ISRS and THz excitation has been demonstrated. Indeed, there exists only one published example of THz excitation and direct optical probing of an underdamped vibrational mode in any material [10].

---

[a] Contributed equally to this work.
[b] Author to whom correspondence should be addressed. Electronic mail: jjohnson@chem.byu.edu

Lithium niobate (LiNbO$_3$), often used for high-field THz generation [11,12], is a trigonal ferrolectric crystal ($T_C$ = ~1210 K) [13]. Strong lattice anharmonicities allow for strong coupling between light and optical phonons, leading to a prototypical phonon-polariton dispersion curve due to light coupling to the soft phonon mode. A number of ISRS studies have been performed on LiNbO$_3$, mainly in an effort to map out the phonon-polariton dispersion [14-17]. A recent study showed radiative control over vibrational modes in lithium niobate using ISRS in an optical Kerr effect measurement scheme to simultaneously excite and probe both forward and backward propagating phonon-polaritons [18].

Here we show a direct comparison of LiNbO$_3$ lattice control via resonant THz excitation versus nonresonant ISRS excitation of phonon-polaritons. We demonstrate selective excitation of the forward propagating phonon-polariton using high-power broadband single-cycle THz pulses; whereas 800 nm ISRS excitation drives both forward and backward propagating phonon-polaritons [18]. The amplitude of the ISRS excited modes saturate as the 800-nm pump fluence is increased. However, under THz excitation, the oscillation amplitude increases linearly with incident THz electric field strength, even becoming superlinear at the highest field-strengths indicative of anharmonic effects (see below for more discussion). We therefore demonstrate advantages of THz excitation including selective excitation of only the forward phonon-polariton with no observed saturation of the mode as the pump field strength is increased. Exciting the phonon-polariton to extreme amplitudes allows for a first of its kind extraction of the shape of the interatomic potential energy surface over the range of atomic motion.

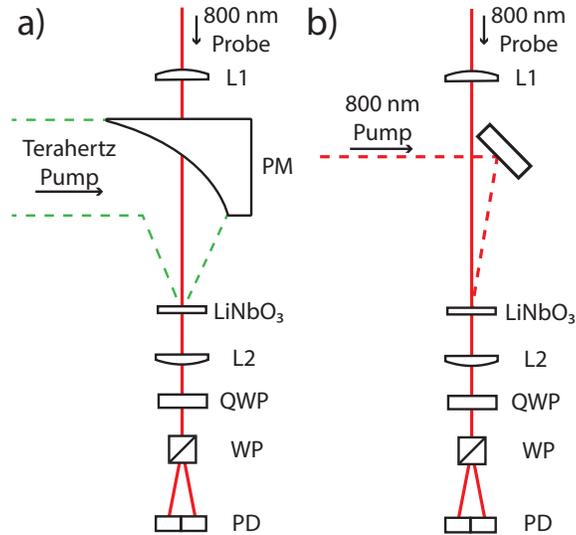

**Figure 1.** Experimental layouts: the probing scheme is identical in each. The probe is focused to the sample position by a lens (L1). After passing through the sample, the beam is collimated using another lens (L2). Polarization changes to probe by sample excitation are converted to intensity changes by the combination of a $\lambda/4$ waveplate (QWP) and a Wollaston prism (WP). The difference in intensity between perpendicular polarizations after the WP are monitored with balanced photodiodes (PD). a) High-field THz radiation is focused to the sample using an off-axis parabolic mirror (PM). b) 800 nm excitation light is directed to the sample for ISRS measurements.

As shown in Fig. 1, the experimental setup used for this study could be altered to excite the sample, a 0.5 mm x-cut $LiNbO_3$ crystal, with either a high-field THz pump pulse or an 800 nm ISRS pump pulse. Both setups used the identical time-delayed 800 nm probing scheme with the probe beam initially polarized parallel to the sample ordinary optical axis. Pump and probe pulses were focused and overlapped spatially on the sample. The (THz or ISRS) excitation induces the lattice vibration, and the probe pulse interrogates the phonon-polariton vibration via a Raman scattering process [18] (this mode is necessarily both IR and Raman active). Polarization changes to the probe beam are recorded by directing the probe beam through a quarter-wave plate (QWP) with fast axis set at 45° to the initial probe polarization, followed by a Wollaston prism (WP) to split it into separate vertically and horizontally polarized beams (identical to a standard electro-optic sampling configuration [19]). Balanced photodiodes (PD) measure the difference in intensity of the two pulses and the resulting signal is recorded digitally [20]. The pump path was mechanically chopped resulting in pump pulses at 500 Hz and probe pulses at 1 kHz repetition rates.

Intense, broadband THz pulses were generated by optical rectification of 1450 nm pulses in the organic crystal OH1 [21,22]. 1450 nm light was generated by optical parametric amplification of 800 nm pulses derived from a Ti:Sapphire laser. The 1450 nm pulses had energy of ~0.9 mJ and a pulse duration of ~100 fs. After THz generation in the OH1 crystal, a Teflon filter was used to block residual IR and visible light. To prevent absorption of the THz by water vapor, the THz beam path from generation to sample position was purged with dry air to a relative humidity of less than 2%. The field strength of the THz incident on the sample was controlled by rotating the first of two wire-grid polarizers while the second remained fixed. THz waveforms were recorded by electro-optic sampling with a GaP crystal at the sample position (100 μm (110) GaP bonded to 1 mm (100) GaP); the peak-to-peak THz field strengths ranged from 0.1 to 1.2 MV/cm. For optimal excitation of the phonon-polariton, the THz was polarized along the $LiNbO_3$ extraordinary axis.

The ISRS measurements utilized loosely focused 800 nm, ~100 fs pump pulses with a $1/e^2$ radius of 1.36 mm at the sample position, polarized at 45° to the sample optical axes. The incident 800 nm pump fluence was controlled with a variable attenuation filter and ranged from 0.17 to 13.9 mJ/cm$^2$. As shown in Fig. 1b, to avoid 800 nm pump light reaching the detectors, the pump and probe pulses crossed the sample at a small angle, allowing spatial filtering of the pump beam.

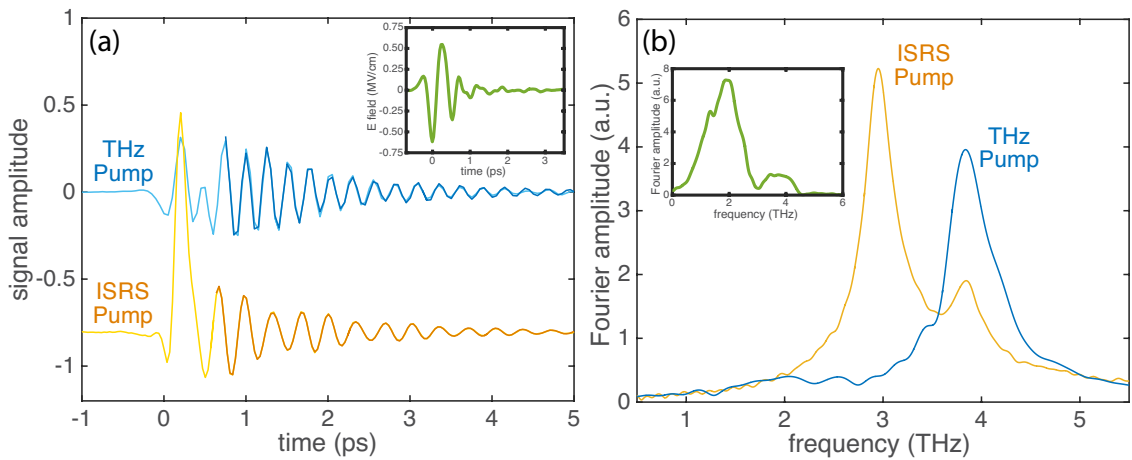

**Figure 2.** a) Time-domain and b) frequency-domain traces of phonon excitation in $LiNbO_3$. a) Light blue is a trace recorded with THz excitation. The light orange trace (that is shifted downwards) was recorded with 800 nm excitation. The dark line overlaying the data is a fit to the data. The inset shows the incident THz electric field. b) The Fourier transform of the decaying oscillations in the time domain traces. ISRS excitation (orange) shows two peaks at 2.9 and 3.9 THz. THz excitation in blue selectively shows one peak at 3.9 THz. The inset shows the spectrum of the incident THz electric field.

Fig. 2a shows time-domain traces of lithium niobate under THz excitation (upper in blue) and ISRS excitation (lower in orange). In both traces, a signal similar to the profile of the excitation pulse is observed at the beginning of the waveform, followed by the damped oscillations of the excited vibrations. In the case of THz excitation, this initial signal is due to atomic motion caused by the incident THz radiation (see the Fig. 2 insets) combined with an electro-optic signal contribution, and in the Raman excitation it is due the instantaneous hyperpolarizability induced by the 800-nm light. The darker solid line over the damped oscillatory signal is from an underdamped harmonic oscillator fit to the data (see below). Fourier transforms of the decaying oscillations are shown in Fig. 2b. Two peaks are observed in the ISRS frequency-domain trace at ~3.9 THz and ~2.9 THz which correspond respectively to excited forward and backward propagating phonon-polaritons [18]. The backward propagating phonon-polariton contributes a larger signal at 2.9 THz, which may obscure the signal from the 3.9 THz mode. In Ref [18], the lower phonon polariton dispersion curve for lithium niobate is calculated and the wavevector matching considerations relevant to our current measurements are outlined. The two relevant wavevector equations are [18]

$$k_+ = \frac{n_o}{c}\omega - \frac{n_e}{c}(\omega - \Omega_+) \tag{1}$$

$$k_- = \frac{n_e}{c}\omega - \frac{n_o}{c}(\omega - \Omega_-), \tag{2}$$

where $k_+$, $\Omega_+$, $k_-$, and $\Omega_-$ are respectively the wavevector and frequency of the forward and backward going phonon-polaritons. $\omega$ is the light frequency, $c$ is the speed of light, and $n_o$ and $n_e$ are the ordinary and extraordinary refractive indices at the laser wavelength. These wavevector selection rules hold for both 800-nm excitation and probing processes, thus only phonon-polaritons with these two wavevectors can be excited and/or probed with 800 nm light. The two wavevectors correspond with the observed 2.9 and 3.9 THz frequencies. In contrast to ISRS excitation, the THz excitation frequency-domain trace shows only a single peak. This indicates that THz excitation selectively excites only the forward propagating phonon-polariton, and the probe wavevector selection for $k_+$ results in only observing an oscillation at 3.9 THz. This can be understood as the phonon-polariton is a coupled mode between the crystal vibration and the THz light propagating in the crystal, and the incident THz is initially only propagating in the forward direction. We likely excite phonon-polaritons with many frequencies and the associated wavevectors traveling in the forward direction due to our broadband THz pump pulse, and yet we selectively observe only the 3.9 THz mode due to the wavevector selectivity of the probing scheme.

As stated above, the probe setup was identical for THz and ISRS measurements. In an effort to directly compare the strength of THz excitation of the 3.9 THz mode to that of ISRS, time-domain traces at each excitation power were fit using a nonlinear fitting routine to a single damped harmonic oscillator in the case of THz excitation, and to two oscillators in the case of Raman excitation. Fig. 3 shows the extracted amplitudes of the 3.9 THz oscillator at each power measurement for both THz and ISRS excitation, scaled proportional to the excitation volume relevant for each measurement. In each case, the optical probe setup is identical, but the signal amplitude is dependent on the excited volume the probe beam interrogates. In ISRS excitation, the excited volume is the sample thickness of 500 μm. On the other hand, the THz radiation is strongly absorbed leading to an excitation depth (and therefore excited volume) of only ~ 60 μm [23]. The bottom-axis shows the 800 nm fluence for the ISRS traces, and the upper-axis shows the peak-to-peak field strength for THz excitation, and the inset is a zoomed-in look at the lower amplitude region. The scale of each horizontal axis was constructed to directly compare amplitudes for both types of excitation in the low-power, linear regime of the power-dependence. As the 800 nm fluence increases, there is notable saturation in the oscillation amplitude of both phonon-polariton modes (see Fig. 3 inset). The dotted orange line in Fig. 3 is a fit to a simple exponential model to allow some extrapolation of the oscillation amplitude saturation to higher fluences. We suspect this saturation is a result of multiple-photon absorption in lithium niobate, followed by free-carrier absorption (see Supplementary Material Online). This would reduce the pump fluence throughout the crystal and could disrupt the excitation or probing process. Importantly in contrast, no saturation is seen in the THz excitation amplitudes; as the THz field strength is increased, the oscillation amplitude increases nearly linearly (with positive deviations at large amplitudes). From this comparison, we note that only a modest THz electric field strength of 50 kV/cm is required to induce atomic motion equivalent to a relatively high 800 nm pump fluence of ~ 5 mJ/cm$^2$. At the maximum electric field strengths used, THz excitation leads to atomic motion 16.5 times larger than the level at which ISRS excitation appears to saturate. Therefore, in addition to the selective excitation of only the forward going phonon-polariton, THz pumping is able to drive this phonon to very large amplitudes without saturation.

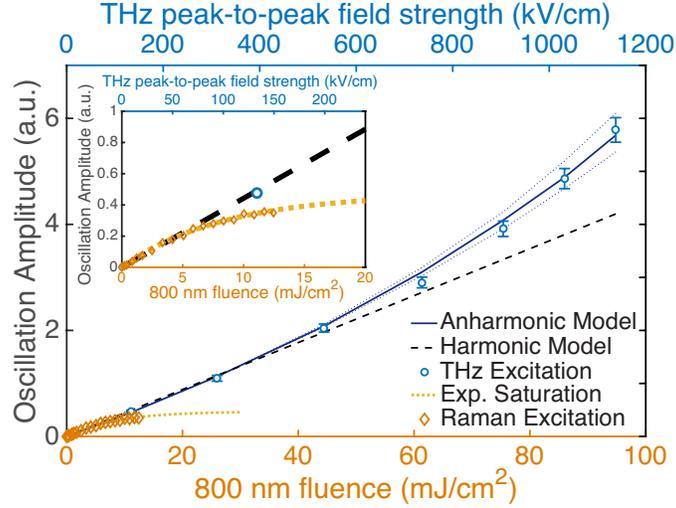

**Figure 3.** Oscillation amplitudes scaled by the probed volume of the 3.9 THz phonon-polaritons under THz excitation (blue circles) and ISRS (orange diamonds) as a function of peak-to-peak electric field strength and fluence respectively. Peak-to-peak THz field strengths are shown in the upper axis and 800 nm fluence in the lower axis. The dashed black line is the expected oscillation amplitude from due to a harmonic potential. The solid blue line is the expected oscillation amplitude due to an anharmonic potential that shows superlinear behavior. The dotted blue lines accompanying the anharmonic model indicate ±15% values for the cubic $\lambda$ parameter in the anharmonic model. The dotted orange line is from a fit to a simple exponential model to extrapolate the ISRS oscillation amplitude saturation to higher fluences. Inset: Expanded view of the low amplitude region.

We now comment on the implications of the observed results with respect to exceedingly large amplitude nuclear motion control in LiNbO$_3$. THz excitation allows driving vibrational motion efficiently to much larger amplitudes than ISRS excitation. With ISRS, nonlinear pump absorption hinders large amplitude motion along the soft mode coordinate (see Supplementary Material Online). It has been suggested that multiple pulse excitation could be utilized to coherently excite the mode to greater amplitudes [24]. In contrast, Refs. [25,26] have suggested that a single transform limited pump pulse, as in our ISRS measurements presented here, is more efficient in exciting to large amplitudes than any kind of pulse-shaped excitation; however, in the theoretical analysis of [26] they neglected nonlinear absorption at high fluences such as that observed here. In any case, it appears THz excitation is preferred to ISRS excitation. The vibrational amplitude at the highest THz electric fields measured is approximately 17 times larger than that reached by ISRS emphasizing clear advantages of THz excitation. Indeed, we observe at the highest peak-to-peak THz field strengths a departure in the oscillation amplitude from linearity, consistent with exciting to oscillation amplitudes that sample anharmonic regions of the potential energy surface; as the mode softens at high amplitudes, the initial distance from equilibrium positions becomes larger than the linear trend a simple harmonic model predicts.

By adding a cubic term to a harmonic potential, and solving the subsequent equations of motion for the resulting driven anharmonic oscillator, we can compare the model to the data and estimate the shape of the real potential energy surface. Incorporating all of our THz pump traces at varying electric field strengths into a fitting routine simultaneously allows us to estimate the amplitude of a cubic anharmonic term in the potential, which we model with the following equation:

$$U(x) = \frac{1}{2}\omega_0^2 x^2 - 3\lambda\omega_0^2 x^3, \qquad (3)$$

where $\omega_0$ = 3.89 THz and $\lambda$ = -0.0533 are the best fit parameters. Results of this modeling and fitting are displayed in Figures 3 and 4. In both figures, the dotted lines accompanying the anharmonic model predictions indicate $\pm15\%$ values of the cubic parameter $\lambda$ to give an indication of the precision of our anharmonic potential determination. We see the superlinear trend observed in the experimental data in Fig. 3 is reproduced quite well, as well as a slight softening of the mode seen in the inset to Fig. 4. It is instructive to note that in this case, the largest departure from the linear harmonic amplitude increase due to anharmonic effects is 38%, whereas the maximum mode softening is only 1.3%. Until this point, anharmonic vibrational effects initiated with THz excitation have only been observed in THz transmission measurements through a field-strength dependent change in observed resonant frequency of the mode of interest [27,28]. Our current results suggest that direct observation of the vibrational motion in the time domain as demonstrated here is a much more sensitive probe of anharmonic vibrational effects. Indeed, we are able to extract the shape of the anharmonic potential over a limited range directly from fits to our experimental data; historically, information about the shape of the interatomic potential in solid materials has only been accessible through first principles calculations. In Fig. 4 we show a comparison between the simple harmonic potential and the anharmonic potential we extract from fits to the data. The dotted vertical lines indicate how far up the potential the oscillation traverses at the highest incident THz field-strength, which also indicates the extent to which accurately extracting the potential should be plausible. With future measurements utilizing larger peak-to-peak driving fields, even more of the anharmonic potential can be examined and the possibility of ultrafast polarization switching explored. Ultrafast x-ray diffraction measurements that can quantitatively determine the motion in real space, such as those in [29,30] would be incredibly useful to confirm the current findings, as well as similar measurements on other materials to test the potentially universal advantages of THz excitation.

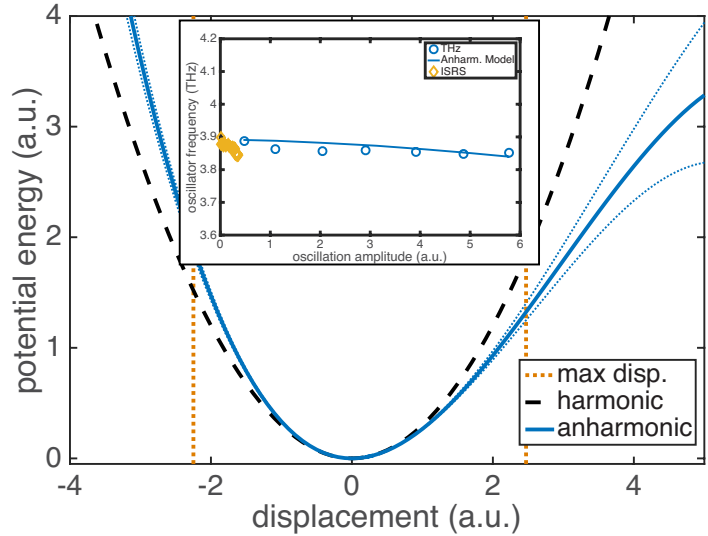

**Figure 4.** Potential energy surface calculated from fits to data of anharmonic potential with harmonic and cubic terms (solid blue line) compared to simple harmonic potential (dashed black line). As in Fig. 3, the dotted blue lines accompanying the anharmonic model indicate ±15% values for the cubic $\lambda$ parameter in the anharmonic model. The vertical orange dotted lines indicate the maximum displacement we reach at the maximum THz electric field strength. Inset: Extracted oscillation frequency as a function of oscillation amplitude (from Fig. 3) for THz excitation (blue circles) and ISRS excitation (orange diamonds). The solid blue line indicates the modeled mode softening using the extracted anharmonic potential.

In summary, we used single-cycle, broadband THz and 800 nm pulses to selectively excite the phonon-polariton(s) in $LiNbO_3$ and measured the resulting dynamics with an 800-nm probe pulse. With THz excitation, we observe a single oscillation at 3.9 THz corresponding to the forward propagating phonon-polariton, compared to 800-nm ISRS excitation, where both the forward and backward propagating phonon-polaritons at 3.9 and 2.9 THz respectively are excited. In addition, the amplitude of the excited oscillation does not saturate with THz excitation, indicating that unwanted nonlinear absorption effects, at the present THz field strengths, are not a complicating and potentially destructive factor for THz lattice control. At the highest incident THz electric field-strengths, the oscillator is driven into the anharmonic regime, allowing us to estimate the shape of the potential energy surface beyond the low amplitude harmonic approximation. These results promise potential routes to lattice control using high-field THz radiation instead of ISRS, including driving oscillations into the anharmonic regime in order to experimentally study the potential energy surface, previously only accessible through computational efforts, and foreshadows the ability to induce phase changes or polarization switching in ferroelectric materials. Future work with even higher THz electric field strengths will allow an experimental study of larger regions of the potential energy surface and anharmonic coupling between degrees of freedom [31], and even lead to polarization switching tied to the LiNbO$_3$ soft mode displacement [1,32].

The authors acknowledge funding and support from the Department of Chemistry and Biochemistry at Brigham Young University.

**Supplementary Material**

In Fig. 3 of the main text we show that as the 800 nm pump fluence increases, there is notable saturation in the oscillation amplitude of the phonon-polariton. We suspect this saturation is a result of multiple-photon absorption in lithium niobate, followed by free-carrier absorption. This would reduce the pump fluence throughout the crystal and could disrupt the excitation or probing process. The band gap in lithium niobate is such that radiation with wavelength longer than ~300 nm is transmitted; therefore, three-photon absorption of 800 nm would be the expected initial route to carrier excitation. We measured the intensity of the transmitted 800 nm light, which reveals an intensity dependent absorption by the sample, supporting the idea that nonlinear interactions take place at high fluences. In Fig. S1, we show the similar saturation of the oscillation amplitude of both forward and backward phonon-polariton oscillations upon ISRS excitation (left *y*-axis), and also a comparison to the measured transmission of the 800 nm pump (right *y*-axis).

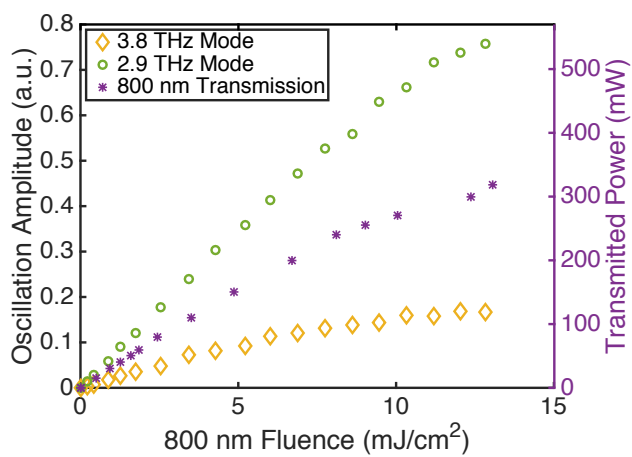

**Figure S1.** Saturation of 2.9 THz (green circles) and 3.8 THz polaritons (orange squares) under ISRS excitation (left axis). Purple stars show the intensity of transmitted 800 nm light (right axis) as a function of incident fluence.

**References**

[1] T. Kampfrath, K. Tanaka, and K. A. Nelson, Nat Photon **7**, 680 (2013).
[2] H. Y. Hwang *et al.*, Journal of Modern Optics **62**, 1447 (2015).
[3] C. H. Matthias and F. József András, Journal of Physics D: Applied Physics **44**, 083001 (2011).
[4] C. Somma, K. Reimann, C. Flytzanis, T. Elsaesser, and M. Woerner, Physical Review Letters **112**, 146602 (2014).
[5] T. E. Stevens, J. Kuhl, and R. Merlin, Physical Review B **65**, 144304 (2002).
[6] H. J. Zeiger, J. Vidal, T. K. Cheng, E. P. Ippen, G. Dresselhaus, and M. S. Dresselhaus, Physical Review B **45**, 768 (1992).
[7] R. Merlin, Solid State Communications **102**, 207 (1997).
[8] Y. X. Yan, E. B. Gamble, and K. A. Nelson, The Journal of Chemical Physics **83**, 5391 (1985).
[9] Y. X. Yan and K. A. Nelson, The Journal of Chemical Physics **87**, 6240 (1987).
[10] T. Huber, M. Ranke, A. Ferrer, L. Huber, and S. L. Johnson, Applied Physics Letters **107**, 091107 (2015).
[11] H. Hirori, A. Doi, F. Blanchard, and K. Tanaka, Applied Physics Letters **98**, 091106 (2011).
[12] K.-L. Yeh, M. C. Hoffmann, J. Hebling, and K. A. Nelson, Applied Physics Letters **90**, 171121 (2007).
[13] R. S. Weis and T. K. Gaylord, Applied Physics A **37**, 191 (1985).
[14] P. C. M. Planken, L. D. Noordam, J. T. M. Kennis, and A. Lagendijk, Physical Review B **45**, 7106 (1992).
[15] H. J. Bakker, S. Hunsche, and H. Kurz, Physical Review B **50**, 914 (1994).
[16] H. J. Bakker, S. Hunsche, and H. Kurz, Reviews of Modern Physics **70**, 523 (1998).
[17] C. C. Lee, C. T. Chia, Y. M. Chang, M. L. Sun, and M. L. Hu, Japanese Journal of Applied Physics **43**, 6829 (2004).
[18] Y. Ikegaya, H. Sakaibara, Y. Minami, I. Katayama, and J. Takeda, Applied Physics Letters **107**, 062901 (2015).
[19] J. A. Johnson, F. D. J. Brunner, S. Grübel, A. Ferrer, S. L. Johnson, and T. Feurer, J. Opt. Soc. Am. B **31**, 1035 (2014).
[20] C. A. Werley, S. M. Teo, and K. A. Nelson, Review of Scientific Instruments **82**, 123108 (2011).
[21] F. D. J. Brunner, O. P. Kwon, S.-J. Kwon, M. Jazbinšek, A. Schneider, and P. Günter, Opt. Express **16**, 16496 (2008).
[22] C. Ruchert, C. Vicario, and C. P. Hauri, Opt. Lett. **37**, 899 (2012).
[23] M. Unferdorben, Z. Szaller, I. Hajdara, J. Hebling, and L. Pálfalvi, Journal of Infrared, Millimeter, and Terahertz Waves **36**, 1203 (2015).
[24] A. M. Weiner, D. E. Leaird, G. P. Wiederrecht, and K. A. Nelson, Science **247**, 1317 (1990).
[25] O. V. Misochko, M. V. Lebedev, H. Schäfer, and T. Dekorsy, Journal of Physics: Condensed Matter **19**, 406220 (2007).

[26]	T. Shimada, C. Frischkorn, M. Wolf, and T. Kampfrath, Journal of Applied Physics **112**, 113103 (2012).
[27]	M. Jewariya, M. Nagai, and K. Tanaka, Physical Review Letters **105**, 203003 (2010).
[28]	I. Katayama *et al.*, Physical Review Letters **108**, 097401 (2012).
[29]	S. Grübel *et al.*, in *ArXiv e-prints* **arXiv:1602.05435** (2016).
[30]	F. Chen *et al.*, Physical Review B **94**, 180104 (2016).
[31]	M. Forst, C. Manzoni, S. Kaiser, Y. Tomioka, Y. Tokura, R. Merlin, and A. Cavalleri, Nat Phys **7**, 854 (2011).
[32]	T. Qi, Y.-H. Shin, K.-L. Yeh, K. A. Nelson, and A. M. Rappe, Physical Review Letters **102**, 247603 (2009).